\PassOptionsToPackage{unicode}{hyperref}
\PassOptionsToPackage{hyphens}{url}
\PassOptionsToPackage{dvipsnames,svgnames,x11names}{xcolor}
\documentclass[
  11pt,
]{article}
\usepackage{xcolor}
\usepackage[margin=1in]{geometry}
\usepackage{amsmath,amssymb}
\usepackage{iftex}
\ifPDFTeX
  \usepackage[T1]{fontenc}
  \usepackage[utf8]{inputenc}
  \usepackage{textcomp} % provide euro and other symbols
\else % if luatex or xetex
  \usepackage{unicode-math} % this also loads fontspec
  \defaultfontfeatures{Scale=MatchLowercase}
  \defaultfontfeatures[\rmfamily]{Ligatures=TeX,Scale=1}
\fi
\usepackage{lmodern}
\usepackage{authblk}

\ifPDFTeX\else
\fi
\IfFileExists{upquote.sty}{\usepackage{upquote}}{}
\IfFileExists{microtype.sty}{% use microtype if available
  \usepackage[]{microtype}
  \UseMicrotypeSet[protrusion]{basicmath} % disable protrusion for tt fonts
}{}
\makeatletter
\@ifundefined{KOMAClassName}{% if non-KOMA class
  \IfFileExists{parskip.sty}{%
    \usepackage{parskip}
  }{% else
    \setlength{\parindent}{0pt}
    \setlength{\parskip}{6pt plus 2pt minus 1pt}}
}{% if KOMA class
  \KOMAoptions{parskip=half}}
\makeatother
\usepackage{longtable,booktabs,array,tabularx}
\usepackage{calc} % for calculating minipage widths
\usepackage{etoolbox}
\makeatletter
\patchcmd\longtable{\par}{\if@noskipsec\mbox{}\fi\par}{}{}
\makeatother
\usepackage{graphicx}
\makeatletter
\def\fps@figure{htbp}
\makeatother
\providecommand{\tightlist}{%
  \setlength{\itemsep}{0pt}\setlength{\parskip}{0pt}}
\usepackage[]{natbib}
\usepackage{bookmark}
\IfFileExists{xurl.sty}{\usepackage{xurl}}{} % add URL line breaks if available
\hypersetup{
  pdftitle={DeltaSelect: Affordable A/B Testing for Coding Agents},
  pdfauthor={Nicholas J. Conn},
  colorlinks=true,
  linkcolor={Maroon},
  filecolor={Maroon},
  citecolor={Blue},
  urlcolor={Blue},
  pdfcreator={LaTeX via pandoc}}

\title{DeltaSelect: Affordable A/B Testing for Coding Agents}
\author{{Nicholas J. Conn, PhD}}
\affil{Conn Castle Studios\\
\href{mailto:nick@conncastle.com}{nick@conncastle.com}}
\date{}

\begin{document}
\maketitle

\begin{abstract}

Coding-agent benchmarks are built for broad and comprehensive
comparisons, not frequent development decisions. Individual runs vary,
full suites are expensive, and the benchmark harness may differ from the
harness used in practice. In a resampling analysis of DeepSWE's published
trials, only 19.5\% of tasks (22 of 113) had a fifth-percentile Pearson
correlation of at least 0.50 with full-benchmark performance. The paper
presents DeltaSelect, an open-source method that identifies tasks whose
one-run results consistently track full-benchmark performance using Pearson
correlation, maps fractional verifier results to a common score using linear
regression, and selects a fixed task set within a dollar budget. DeltaSelect
is intended for repeated
baseline-versus-candidate comparisons during development, not model
rankings. In a gpt-5.6-luna low-reasoning case study, DeltaSelect was used to revise
custom skills and instructions. Across 13 evaluations, the recorded
cost was \$27.86 at rates published August 16, 2026. The
adopted version cost 58.1\% less than the initial version
(\$1.75 versus \$4.18; \emph{p}=0.008), while the calibrated
score was higher (42.36\% versus 36.46\%; published-analog
variance \emph{p}=0.326).

\end{abstract}

\section{Introduction}\label{introduction}

Developers routinely change skills and instructions, tune reasoning
levels, switch models, and make other changes intended to improve their
coding agents. Improvement may mean a higher score, a lower cost, fewer
failures, or a useful tradeoff between these goals. Regardless of the goal,
iteration is central to research and development. Comparisons must be low
cost enough to repeat throughout the development process while providing
enough statistical power to distinguish meaningful changes from
run-to-run variation.

Full benchmarks remain the right tool for broad comparisons, but their
cost prevents routine use during development. This paper presents
DeltaSelect, a method that uses repeated published trials to select a
small, fixed task set within a specified budget and compare a baseline
with a candidate. DeltaSelect is not a leaderboard, a universal model
ranking, or a replacement for a full benchmark. The research question is
specific: \emph{Can a single run per selected task provide useful evidence
for frequent baseline-versus-candidate comparisons without full-benchmark
cost?}

One barrier to low-cost comparison is how benchmark results are typically scored.
Binary scoring discards useful information from each run. DeepSWE
executes each task four times and scores each run as pass or fail
\citep{huang2026deepswe}. Repeating each task four times reduces the
effect of run-to-run variation, but does not recover discarded partial
progress and increases the benchmark's scored-run cost four-fold.

DeepSWE's released verifier records contain more information. In
addition to binary resolution, the records report the fraction of new
fail-to-pass tests passed (F2P) and the fraction of existing
pass-to-pass tests preserved (P2P). F2P distinguishes partial
progress mapped to zero by binary scoring. F2P varies across runs and is
not equally informative for every task. F2P also measures verifier tests,
not the percentage of a task's requirements completed: tests can overlap,
differ in difficulty, or encode several assertions. Even with those
limits, F2P preserves partial progress that binary scoring discards. An
all-or-nothing result is too coarse for practical one-run comparisons.

\begin{figure}
\centering
\includegraphics[width=\linewidth,keepaspectratio,alt={Ranked plot of 113 tasks showing fifth-percentile correlation, median correlation, and the fifth-to-95th-percentile resampling band.}]{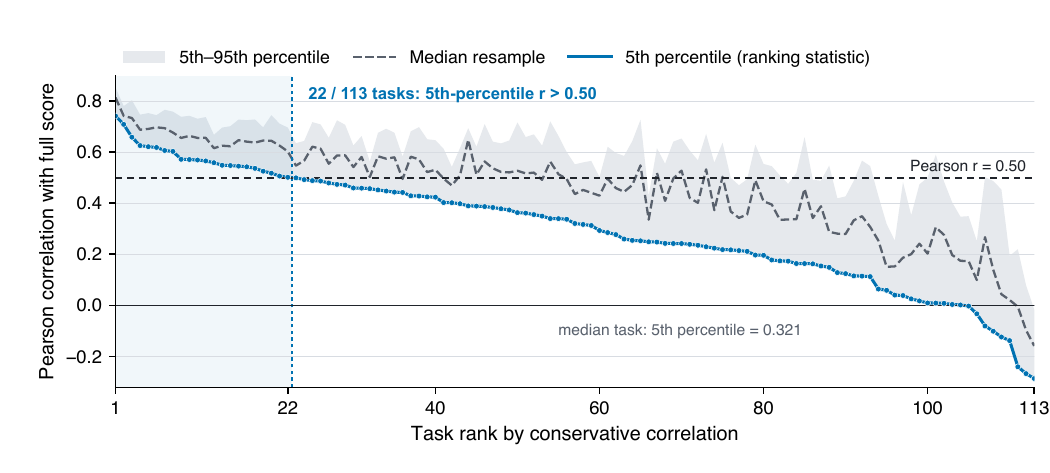}
\caption{Which tasks remain useful when run only once? For each task,
one published run per model--effort configuration is resampled 10,000
times and correlated with the full-benchmark score. Blue points and the
solid line show the fifth percentile, the dashed line shows the median,
and the shaded band spans the fifth--95th percentiles. Only 22 of 113
tasks (19.5\%) exceed 0.50 on the conservative ranking statistic; most
tasks are weak one-run proxies for the observed full-score variation.}
\label{fig:task-reliability}
\end{figure}

Reducing cost to one run per task then creates a task-selection problem.
DeltaSelect ranks tasks by how consistently one published run correlates
with full-benchmark performance across the observed configurations. The
ranking does not judge a task's broader benchmark value; it measures
whether one run provides useful evidence for the development decision
studied here. A low one-run correlation does not invalidate a task or
diminish its value in a comprehensive benchmark. It means only that a
single execution provides weak evidence for this development decision.
Across 10,000 deterministic one-trial resamples, the
median task's fifth-percentile Pearson correlation with the full DeepSWE
configuration score was 0.321. Only 22 of 113 tasks reached 0.50, and
only two exceeded 0.70. Reliable one-run signal was limited to a small
part of the benchmark (Figure~\ref{fig:task-reliability}).

A second challenge with many leading benchmarks is the harness gap. A
coding-agent score reflects the entire system used to produce it: the
model, reasoning setting, client, tools, instructions, and execution
harness. A published benchmark may deliberately fix a neutral harness to
isolate other differences. A developer instead cares about the harness
that will be used in practice, which may have different context
management, tools, prompts, and control flow. On Terminal-Bench 2.1,
holding model and reasoning effort fixed while changing the harness moved
GPT-5.5 xhigh by 5.1 percentage points and Opus 4.7 max by 2.8 points
\citep{terminalbench2026}. Changing the harness can therefore move a
score even when the model and reasoning level do not change. Selected
tasks must be validated in the intended harness before use.

Using frontier models creates another constraint: a comparison needs
headroom. Frontier agents are beginning to saturate established coding
benchmarks, making further improvements harder to distinguish
\citep{openai2026verified}. A baseline near 100\% on every selected task
leaves little room for a better candidate to raise the score. A baseline
near 0\% leaves no room to detect a further regression. Selecting
tasks with usable headroom can mitigate the problem, but
the mitigation is not free: those tasks may be less repeatable or more
expensive. The choice is a tradeoff between headroom, repeatability, and
price. Published results are used to build the task set and scoring model.
The baseline must then be run in the exact harness and version
intended for the comparison. That step cannot be skipped: changing or
updating the harness can change task behavior, score, and cost. Only that
target-harness baseline can establish the selected tasks' headroom, repeatability,
and starting score.

\begin{figure}
\centering
\includegraphics[width=\linewidth,keepaspectratio,alt={Four-panel workflow showing one published run sampled per configuration, repeated samples forming a correlation distribution, tasks ranked by the fifth percentile, and a fixed ranking scanned under a dollar budget.}]{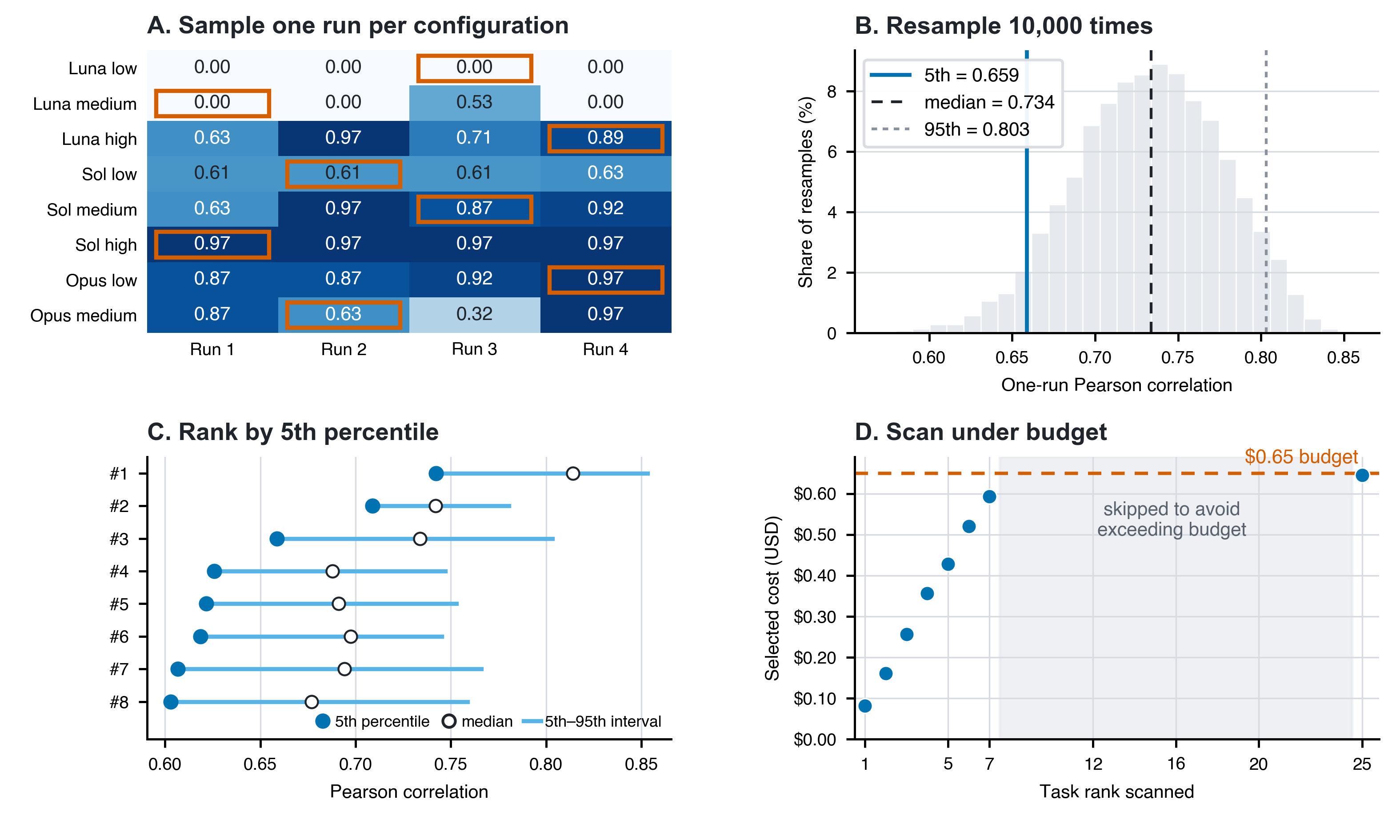}
\caption{Selecting a fixed task set under a dollar budget. (A) One
completed run is sampled per model--effort configuration for each task;
eight configurations are shown for the DeepSWE task
\texttt{koota-pair-relation-tracking}, which ranks third.
Sol denotes gpt-5.6-sol, and Opus denotes Claude Opus 4.8.
(B) Repeating the sampling 10,000 times produces each task's correlation
distribution. (C) Tasks are ranked by the fifth percentile of task-level
distributions. (D) The fixed ranking is scanned under the execution
budget using snapshot prices: ranks 1--7 fit, ranks 8--24 exceed the
remaining \$0.056, and
rank 25 is the next affordable task. Task rank is shown horizontally so
the skipped interval is explicit.}
\label{fig:selection-pipeline}
\end{figure}

DeltaSelect addresses this decision with a deliberately simple pipeline.
The pipeline ranks tasks by fifth-percentile correlation across repeated
public trials, maps each task's F2P result onto the published full-score
scale, weights tasks with smaller calibration errors more heavily, and
selects the highest-ranked tasks that fit the model-specific dollar
budget. Figure~\ref{fig:selection-pipeline} shows these steps. The
resulting tasks, repetitions, calibration coefficients, weights, price
assumptions, snapshot checksum, and seeds are frozen before the first
baseline-versus-candidate comparison and remain fixed throughout
iteration.

The method is intended to compare two versions of the same coding-agent
system under a fixed specification. DeltaSelect is not a leaderboard, a universal model
ranking, proof of unseen-task generalization, or a replacement for
comprehensive benchmarks. A small correlated task set does not inherit
the full benchmark's breadth merely by tracking the historical full score.

The paper makes three contributions:

\begin{enumerate}
\def\labelenumi{\arabic{enumi}.}
\tightlist
\item
  A deterministic task ranking based on repeated trials to account for
  run-to-run variation instead of ranking tasks from mean results alone.
\item
  A calibrated score retaining fractional verifier evidence
  discarded by binary scoring while keeping raw F2P, P2P, and binary
  outcomes visible.
\item
  A model- and effort-specific budget scan based on observed task
  prices without allowing price to change the reliability ranking.
\end{enumerate}

The central claim is intentionally narrow: repeated public trials can identify
a small, fixed set of reliable tasks for affordable baseline-versus-candidate
comparisons. In the published example, the selected tasks are projected to
distinguish two materially different model--effort configurations at far lower
scored-trial cost than the full benchmark. Target-harness runs show why the selected tasks
must then be validated and baselined in the intended harness. The case study shows how a
frozen task set can guide development toward higher score, lower cost, or both.

\section{Related Work}\label{related-work}

\subsection{Data-efficient model
evaluation}\label{data-efficient-model-evaluation}

Prior work provides several techniques for estimating model performance
with fewer evaluations. The Active Testing technique uses adaptive
sampling and importance weighting to estimate performance with fewer
labels \citep{kossen2021active}. Active Surrogate Estimators adds
learned surrogate predictions to the selection process
\citep{kossen2022surrogate}. Efficient Benchmarking allocates evaluation
resources according to decision reliability and reports large reductions
in HELM evaluation cost with limited reliability loss
\citep{perlitz2024efficient}. These methods reduce evaluation cost by
evaluating a selected subset, allocating runs where they are most informative,
or using surrogate predictions. If examples are sampled at different rates,
importance weighting gives each observed result a compensating influence so
that the estimate represents the original evaluation set rather than the
biased sample.

Short-form benchmark construction provides an even closer precedent.
tinyBenchmarks combines item-response models with anchor examples to
estimate full-benchmark scores from small subsets
\citep{maiapolo2024tiny}. Efficient Multi-Prompt Evaluation treats
prompt choice as another measurement dimension
\citep{maiapolo2024multiprompt}. BenTo reduces benchmarks at the task
level using transferability and facility-location selection, explicitly
considering redundancy \citep{zhao2025bento}.

SubLIME is the closest correlation-based analogue. SubLIME learns to select
1--20\% of benchmark items preserving global rankings, using item
features and partial evaluations from anchor models
\citep{saranathan2025sublime}. SubLIME aims to preserve rankings across
many models. DeltaSelect instead asks whether a change improved or worsened
the same coding-agent system during development. DeltaSelect directly models uncertainty from repeated task
executions, incorporates different dollar costs by task, maps fractional
verifier results to a common reference scale, requires a target-harness
baseline, and freezes a user-visible task set for longitudinal
comparison. A task can help preserve the ranking of many models yet still
be too noisy to compare two system versions from one run.

What is new is the combination for a specific use case: frequent A/B
testing of a coding-agent system during development. DeltaSelect uses
repeated trials to rank task reliability, converts partial-credit results
onto one score scale, selects tasks within a dollar budget, validates them
in the target harness, and reports workflow cost. The task set and scoring
rules then stay fixed while the system changes.

\subsection{Variance and score
validity}\label{variance-and-score-validity}

DeltaSelect measures only run-to-run randomness by repeatedly sampling one of
each task's four published outcomes; it does not measure variation caused by
data, training, or implementation differences
\citep{bouthillier2021variance}.

The method is not specific to DeepSWE. Any benchmark can serve as input if it
reports a full score across configurations and repeated task-level trials for
estimating uncertainty. Terminal-Bench 2.1 is one example: its leaderboard
requires at least five public trials per task and reports an aggregate
task-resolution score \citep{terminalbench2026submission}.

The calibrated score has one intended use: compare a baseline and candidate
on the fixed selected tasks. It should not be read as a general measure of
software-engineering ability. Even when the subset tracks DeepSWE, it does
not cover every capability tested by the full benchmark
\citep{messick1995validity}.

\subsection{Coding-agent benchmarks, harnesses, and verifier
evidence}\label{coding-agent-benchmarks-harnesses-and-verifier-evidence}

SWE-bench evaluates whether an agent can fix real issues in software
repositories \citep{jimenez2024swebench}. SWE-agent showed that
the agent-computer interface materially affects performance
\citep{yang2024sweagent}. Agentless showed that a simpler workflow could
be competitive and inexpensive, reinforcing the need to evaluate cost
and outcome together \citep{xia2025agentless}. Harness-Bench adds recent
controlled evidence that harness choices affect completion, cost,
safety, and process outcomes \citep{yao2026harnessbench}. The current
Terminal-Bench leaderboard provides contemporary same-model comparisons
under native and neutral harnesses, although it offers first-party benchmark
evidence rather than peer-reviewed causal evidence.

DeepSWE provides the repeated long-horizon task matrix used here. DeepSWE
uses original tasks, a fixed mini-swe-agent harness, binary rewards, and
four trials per task--configuration cell; the DeepSWE paper also acknowledges the lack
of partial credit and the tradeoff between harness standardization and
native-product realism \citep{huang2026deepswe}. SkillsBench evaluates
skills across model--harness configurations with repeated task-macro
pass rates and observes harness-mediated effects
\citep{li2026skillsbench}.

F2P is useful, but its numeric detail can imply more certainty than the
verifier tests support. A fraction of F2P tests passed can distinguish a
near-complete implementation from an attempt without meaningful progress. F2P does not
establish that each
test represents an equal, independent unit of the requirement.
Mutation-testing research shows why suite adequacy matters: passing a
weak suite can leave faults undetected \citep{jia2011mutation}.
DeltaSelect therefore preserves raw F2P, P2P, binary resolution, and
build or environment failures beside the aggregate score.

\subsection{Cost-aware agent
evaluation}\label{cost-aware-agent-evaluation}

Prior work also treats score and cost together. AI Agents That Matter argues
for cost-controlled comparisons and accuracy--cost Pareto frontiers, and
shows that repeated sampling can confound comparisons among agent
architectures \citep{kapoor2025agents}. DeltaSelect does not introduce
the idea that agent score and cost should be evaluated jointly. DeltaSelect
addresses measurement expense and attributes realized
spend to the workflow being changed.

DeltaSelect reports score and recorded model cost for every completed
evaluation. A lower cost can be the desired improvement; showing both values
lets the caller decide whether a score change, a cost change, or both justify
adopting a candidate.

\section{DeltaSelect}\label{deltaselect}

DeltaSelect converts repeated trials from an existing benchmark into a
fixed, budget-constrained task set for measuring changes between agent
configurations. The method separates four easily conflated operations:
task ranking, score construction, budget allocation, and
target-harness validation. Ranking asks which tasks remain informative
when run once. Score construction maps heterogeneous task fractions onto
a common reference scale. Allocation asks which ranked tasks fit the
execution budget. Target-harness validation runs the selected tasks in the intended
harness to establish their actual baseline behavior, cost, and headroom before
comparing a candidate.

\subsection{Inputs and notation}\label{inputs-and-notation}

The input is a pinned benchmark snapshot containing repeated fractional
F2P outcomes, full-benchmark scores for observed model--effort
configurations, and trial prices. The analysis uses the DeepSWE v1.1 trial
snapshot \citep{huang2026deepswe}. The snapshot contains 22,586 source trials,
including 22,417 usable trials,
spanning 18 models, 50 model--effort configurations, and 113 tasks. The
169 excluded trials are not silently filled or imputed.

\emph{Luna low}, \emph{Luna medium}, and \emph{Luna high} refer to
gpt-5.6-luna at low, medium, and high reasoning, respectively.

Selection and published-data feasibility examples use the snapshot trial
prices as recorded. Only the target-harness and case-study costs from
Section~\ref{agent-layer-case-study} through
Section~\ref{where-workflow-cost-accumulated}, including the published
estimates compared there, are priced using the August 16, 2026 schedule. Under
that schedule, the \$0.65 published Luna low budget corresponds to
\$0.13.

Let \(t\) index tasks, \(c\) index model--effort configurations, and
\(j\in\{1,\ldots,4\}\) index repeated published runs. Let \(x_{tcj}\) be
the F2P result for run \(j\), and let \(Y_c\) be the fixed
full DeepSWE score for configuration \(c\), calculated as the mean
published score across included trials. \(C_t\) denotes the set of
configurations with a complete four-run cell for task \(t\). The
number of complete configurations can differ by task. Incomplete
configuration--task cells are excluded from the corresponding task's ranking and
calibration rather than reconstructed.

F2P was chosen on evidence. Under the same
10,000-resample procedure, F2P produced 22 tasks with a fifth-percentile
correlation of at least 0.50. All verifier tests combined produced 15,
binary pass/fail produced 11, and P2P alone produced zero. F2P is used
to measure implementation progress; P2P remains a separate regression
check. The choice does not make F2P a literal measure of how much of the
task was completed. The 0.50 threshold is a descriptive reporting cut,
not a selection rule; allocation uses the complete task ranking.

\subsection{Conservative correlation
ranking}\label{conservative-correlation-ranking}

For every task and resample, DeltaSelect independently draws a single trial from the
four complete trials for each available configuration and correlates the
resulting one-run task vector with the fixed full-score vector:

\[
r_t^{(b)} = \operatorname{corr}\!\left(\left[x_{tcJ_{tc}^{(b)}}\right]_{c\in C_t},\;[Y_c]_{c\in C_t}\right), \qquad J_{tc}^{(b)} \sim \operatorname{Uniform}\{1,\ldots,4\}.
\]

The Pearson correlation is recomputed for 10,000 deterministic
resamples. The choice of 10,000 is not statistically special or tuned to the
result. It is a practical large sample that makes Monte Carlo noise small while
adding negligible computation to the task-selection stage. Each pseudorandom task stream is seeded from SHA-256 of the
pinned snapshot checksum, a null byte, and the task identifier, using
the first little-endian 32-bit integer. Constant sampled vectors receive
correlation zero. The ranking statistic is the fifth percentile
of the task-level resampled correlation distribution:

\[
q_t = Q_{0.05}\!\left(r_t^{(1)},\ldots,r_t^{(10{,}000)}\right).
\]

Tasks are ordered from highest to lowest \(q_t\), with ties broken by
median correlation and then task identifier. The fifth percentile is
conservative in a specific sense: a task ranks highly only if most
one-run combinations preserve a strong relationship with full-score
variation across the observed configurations. The statistic is not a 90\%
confidence bound on future models, tasks, or harnesses. The statistic is
conditional on the four retained trials and the finite configuration
matrix.

The selector does not choose tasks that best separate one favorable historical
pair. It asks whether tasks continue to track configuration differences after
each four-run cell is reduced to one sampled outcome.

\subsection{Per-task calibration and
weighting}\label{per-task-calibration-and-weighting}

Task fractions are not directly comparable. A 0.50 F2P result on one
repository can represent a different level of observed system
performance than 0.50 on a second repository. DeltaSelect fits a separate ordinary
least-squares mapping from each task's four-run mean to the common
published full-score scale:

\[
Y_c = \alpha_t + \beta_t\,\bar{x}_{tc} + \varepsilon_{tc}.
\]

\begin{figure}
\centering
\includegraphics[width=\linewidth,keepaspectratio,alt={Two-panel score-construction workflow showing per-task ordinary least-squares mappings and normalized inverse residual-variance weighting.}]{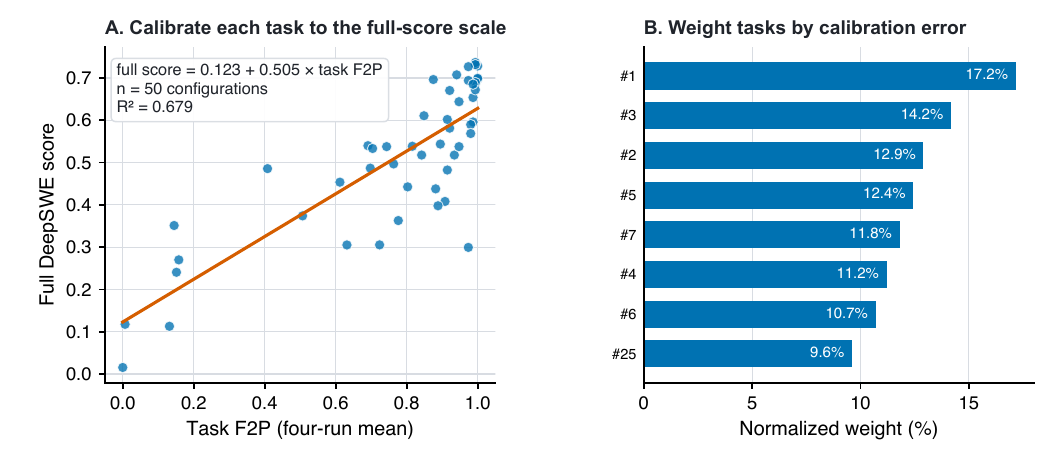}
\caption{Score construction is separate from task selection. (A) For
each selected task, ordinary least squares maps F2P outcomes
onto the common full-score scale; the DeepSWE task
\texttt{koota-pair-relation-tracking} is shown.
(B) Normalized inverse residual-variance
weights combine the calibrated task estimates; rank labels correspond
to Figure~\ref{fig:selection-pipeline}. The operations affect the aggregate score but do not alter
task ranking or budget allocation.}
\label{fig:score-construction}
\end{figure}

Here \(\bar{x}_{tc}\) is the mean of the four published F2P trials. The
fitted line is later applied to the noisier one-run observation in a
low-cost evaluation. Ordinary least squares was chosen for transparency
and auditability, not because the relationship is known to be linear.
This mapping is also the normalization step. Raw F2P fractions come from
different verifiers and do not carry the same meaning across tasks. After
calibration, every task estimates the same full-score quantity. The normalized
weights sum to one, so the aggregate remains on that common scale rather than
growing with the number of selected tasks. A task's number of verifier checks
affects the granularity of its F2P fraction, not the scale of the aggregate.
Calibration does not remove dependence between related tasks.
The example calibration curve (Figure~\ref{fig:score-construction}) shows
possible structure near high F2P values; nonlinear alternatives remain
unmeasured future work.

Residual error determines task influence in the final selected set
\(S\):

\[
\hat{\sigma}_t^2 = \frac{1}{|C_t|-2}\sum_{c\in C_t}\left(Y_c-\hat{Y}_{tc}\right)^2, \qquad w_t = \frac{\hat{\sigma}_t^{-2}}{\sum_{j\in S}\hat{\sigma}_j^{-2}}.
\]

Tasks with undefined or nonpositive residual variance are excluded to
avoid undefined infinite weights. The remaining
positive inverse-error weights are normalized over the final selected
set. A task with smaller linear-calibration residuals receives
more influence. The weights resemble precision weighting, but the
interpretation must stay limited. Task estimates are correlated, and
each residual variance is estimated from the same finite matrix. The
weights also omit the one-run sampling variance of \(x_t\), a source not
captured by the calibration residual. The weights do not form an optimal
inverse-variance meta-analysis.

For an evaluated arm, the calibrated score is

\[
\widehat{\Theta}(S) = \sum_{t\in S} w_t\left(\hat{\alpha}_t+\hat{\beta}_t x_t\right),
\]

where \(x_t\) is the arm's observed F2P fraction, averaged over
\(m\) repetitions when \(m>1\).

The calibrated score expresses a task's F2P result on the published DeepSWE
full-score scale. This puts otherwise incomparable task outcomes on one common
scale so they can be combined into a single score. It is not an official
DeepSWE score or a literal percentage of requirements completed. DeltaSelect
does not cap the estimate at 0 or 100\%. A linear fit can produce values outside
that range, especially when a new result falls beyond the published
observations. Capping those values would hide the magnitude of changes near the
boundary and alter differences between the baseline and candidate. Values
outside 0--100\% should therefore be reported as model estimates and
interpreted with raw F2P, P2P, binary outcomes, and task failures.

\subsection{Budget allocation}\label{budget-allocation}

Let \(e\) denote the execution configuration used for pricing, including
model and reasoning effort; let \(\bar{c}_{t,e}\) be the observed mean
price of task \(t\) under that configuration; let \(B\) be the per-arm
dollar budget; and let \(m\) be the uniform repetition count.
DeltaSelect scans the fixed reliability order once:

\[
S(B,m,e) = \operatorname{GreedyScan}_{q_t\downarrow} \!\left\{t: m\bar{c}_{t,e} \le B-\sum_{j\in S}m\bar{c}_{j,e} \right\}.
\]

DeltaSelect includes every task fitting the remaining budget. After an
unaffordable task, the scan continues and allows a cheaper lower-ranked task to use the
remaining amount. Figure~\ref{fig:selection-pipeline} shows the resulting
noncontiguous selection for the \$0.65 published Luna low budget: ranks 1--7 fit, ranks 8--24 do
not fit the remaining \$0.056, and rank 25 is the next affordable task.

Neither price nor budget changes the task ranking. Reliability is calculated
before either is considered. Price, budget, model, reasoning effort, and
repetition count affect only which tasks from that fixed ranking are selected;
they do not change the order or task-specific score mapping. Weights are
renormalized over the selected set. The rule is deterministic and easy to
inspect, but it does not guarantee the best possible combination of tasks for
the budget. The scan can skip a combination with better joint information,
choose redundant tasks, or spend less than the budget. The limitations
are accepted in exchange for a simple reproducible specification;
matched-cost comparisons with random, cheapest-first, and
median-correlation selectors have not been run.

The published price schedule is a planning input, not a target-harness-cost
forecast. A mini-swe-agent task price cannot predict the number of
delegated sessions or continuations created by a target-harness multi-agent
workflow. The target-harness baseline must be run before an operating budget is
trusted. The case study in
Section~\ref{the-target-harness-changed-the-measurement} shows how to plan
around this gap: run the intended baseline, check headroom, and compare the
intended reasoning level with the next level up before investing in workflow
optimization.

\subsection{Repeatability ranges}\label{conditional-repeatability-ranges}

For displayed published baseline or comparison points, DeltaSelect
resamples task outcomes and prices 1,000 times, recomputes the selected-set
score and total, and reports the 5th-to-95th-percentile range:

\[
I_{0.90}(Z) = \left[ Q_{0.05}\!\left(Z^{(1:1000)}\right), Q_{0.95}\!\left(Z^{(1:1000)}\right) \right].
\]

This range answers a limited question: if the same published configuration
were run again on the same selected tasks, how much might its reported score or
price move? It does not predict results for new model families, harnesses,
environments, or provider behavior absent from the snapshot.

For a one-run A/B result, DeltaSelect also estimates how far two repeats of the
same unchanged configuration could differ. For each selected task, two
published outcomes from the same configuration are independently sampled with
replacement and transformed into a calibrated difference:

\[
D_0^{(b)} = \sum_{t\in S}w_t\hat{\beta}_t \left(x_{t,J_1^{(b)}}-x_{t,J_2^{(b)}}\right), \qquad R_{0.90} = \left[Q_{0.05}(D_0),Q_{0.95}(D_0)\right].
\]

The observed baseline-minus-candidate difference is reported together with
this range. If the difference falls inside the range, it is consistent with
the repeat-run variation in the published data. If it falls outside, it exceeds
the central 90\% of that variation. This is not a significance test for a new
harness because the candidate was run only once and may have different
variance. When both versions have repeated published runs, the test in the
next subsection provides the formal comparison.

\subsection{Retrospective comparison of two published
arms}\label{retrospective-comparison-of-two-published-arms}

When both arms have repeated published trials, the calibrated weighted
difference and propagated standard error are

\[
\hat{\Delta} = \sum_{t\in S}w_t\hat{\beta}_t (\bar{x}_{tA}-\bar{x}_{tB}), \qquad \operatorname{SE}(\hat{\Delta}) = \sqrt{ \sum_{t\in S}(w_t\hat{\beta}_t)^2 \left( \frac{s_{tA}^2}{n_{tA}} + \frac{s_{tB}^2}{n_{tB}} \right) }.
\]

The two-sided comparison uses the Welch--Satterthwaite approximation for the
effective degrees of freedom \citep{welch1947,satterthwaite1946}. The
calculation assumes separately executed task cells are independent and treats
the selected calibration coefficients and weights as fixed. Related tasks may
share errors, which would violate that independence assumption. The resulting
\emph{p}-value measures how strongly the two published arms differ in this
fixed dataset under those assumptions. The comparison is retrospective because
it was chosen after the data were available rather than preregistered.

\subsection{Freeze, execute, and account for
cost}\label{freeze-execute-and-account-for-cost}

The planner exports the ranked tasks, selected repetitions, calibration
coefficients, weights, price assumptions, source checksum, and random
seeds as a single experiment specification. The same specification is used
for baseline and candidate. The harness records task checksums, client
and model versions, reasoning effort, environment identity, raw F2P and
P2P results, binary resolution, build or verifier failures, calibrated
score, and cost telemetry.

In the Agent Layer case study (Section~\ref{agent-layer-case-study}), every
revision used the same tasks, calibration, weights, and repetition count. This
kept the benchmark itself from changing while the workflow changed. A change
to the tasks or score calculation would define a different benchmark, so its
results would not be combined with the original series.

The frozen specification also defines how cost is counted. It records
request-level token and billing data for model calls caused by the experiment,
including failed, retried, delegated, continued, and abandoned work.

\subsection{Published example: cost to reach statistical
significance}\label{evidence-per-dollar-in-published-trials}

A concrete published-data example tests whether DeltaSelect can recover
a difference that the full DeepSWE matrix already shows clearly. Claude Opus 5
medium scores 20.3 points higher than Gemini 3.6 Flash high on the full binary
score. The pair was chosen because the full benchmark shows a clear difference;
if a small task set cannot distinguish these configurations, it is not ready
for smaller development decisions. The released DeepSWE v1.1 trial artifact
\citep{huang2026deepswe} contains 447 scored Claude trials and 451 scored
Gemini trials, or 898 total. It also records five additional Claude trials that
were excluded from the published score because the agent or verifier errored.
One Gemini task contains only three trials, and DeepSWE does not explain why
the fourth trial is missing. A complete 113-task, four-run protocol for both
configurations would contain 904 trials. The 898 scored trials cost
\$3,063.63.

\begin{figure}
\centering
\includegraphics[width=\linewidth,keepaspectratio,alt={Log-cost comparison of the first DeltaSelect ranked prefix below p equals 0.05, all complete paired tasks, and full DeepSWE.}]{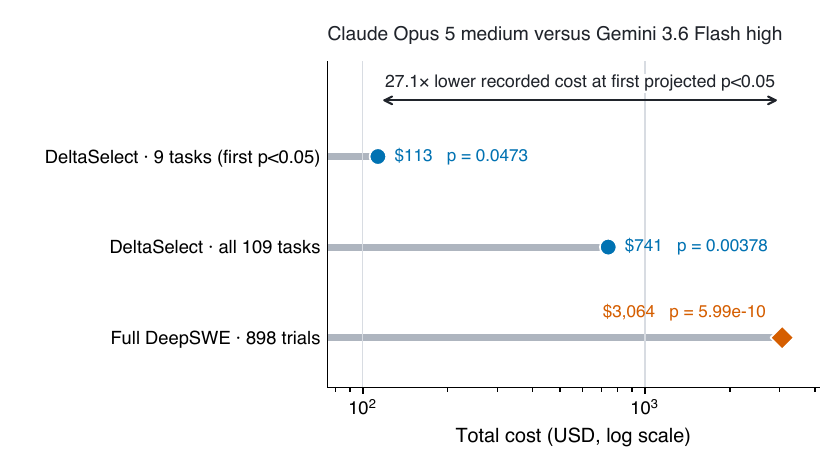}
\caption{Cost to detect a known difference in published data. For Claude Opus
5 medium versus Gemini 3.6 Flash high, the first ranked prefix with projected
\(p<0.05\) contains nine tasks and costs \$113.08 across both configurations
(\(p=0.0473\)). The complete one-run design uses 109 paired tasks and costs
\$740.58 (\(p=0.00378\)); the full published protocol contains 898 trials and
costs \$3,063.63. The projections combine the published four-run mean effect
with one-run task variances and costs. They show expected detectability, not a
realized one-run test.}
\label{fig:evidence-cost}
\end{figure}

The analysis begins with the highest-ranked task and extends the set by one
task at each step, following the fixed ranking. The test from the preceding subsection uses
the published four-run mean difference as the effect and the published one-run
task variances as the expected noise. The result is a projection of a one-run
experiment, not a new experiment that was run. The first prefix with projected
\(p<0.05\) contains nine tasks and costs \$113.08 across both configurations
(\(p=0.0473\)). The nine-task design costs 27.1 times less than the recorded
cost of the full protocol. All
109 tasks with complete cells for both configurations cost
\$740.58 and produce projected \(p=0.00378\). Figure~\ref{fig:evidence-cost}
compares the cost and projected evidence.

The cost difference comes from three changes at once: fewer tasks, F2P instead
of pass/fail, and one run instead of four. This example shows that the complete
DeltaSelect design can detect a large known difference for less projected cost;
it does not show which component creates the saving or what a small instruction
change would cost to detect.

\section{Experiments}\label{experimental-design}

The paper reports two experiments in the Codex harness used by Agent Layer.
First, gpt-5.6-luna runs the eight selected tasks at low, medium, and high reasoning
to test whether published scores, prices, and headroom carry over from
mini-swe-agent to Codex. Second, those same eight tasks are used across 13
evaluations while the custom skills and instructions bundled with Agent Layer,
the author's open-source orchestration layer, are revised
\citep{conn2026agentlayer}. Unless stated otherwise, every Codex result
contains one run per task.

\subsection{Harness validation}\label{target-harness-validation}

Before interpreting a candidate change, the baseline must be run on the
selected tasks in the Codex harness used for the comparison. The validation records client
version, model, reasoning effort, task checksum, environment checksum,
raw and calibrated scores, cost bounds, and task-level ceilings or
floors. Published headroom says which tasks are worth trying; the Codex
baseline determines whether they actually have room to improve.

The eight-task harness comparison uses the set selected by a \$0.65
one-run budget under published gpt-5.6-luna at low reasoning:

{\small
\begin{center}
\begin{tabular}{@{}p{0.48\linewidth}p{0.48\linewidth}@{}}
$\bullet$~\path{koota-composite-trait-aspects} &
$\bullet$~\path{wasmi-trap-coredumps} \\
$\bullet$~\path{koota-pair-relation-tracking} &
$\bullet$~\path{go-git-worktree-merge-conflicts} \\
$\bullet$~\path{testem-bail-on-test-failure} &
$\bullet$~\path{scriggo-method-declarations} \\
$\bullet$~\path{expr-try-catch-errors} &
$\bullet$~\path{bandit-interprocedural-taint-checks}
\end{tabular}
\end{center}
}
The tasks are listed in rank order: the first seven occupy ranks 1--7, and
\path{bandit-interprocedural-taint-checks} occupies rank 25.

gpt-5.6-luna and the same tasks are compared under published mini-swe-agent
and Codex at low, medium, and high reasoning. At low and medium reasoning, a
task-environment checksum records exactly which task revisions Codex ran and
guards against unnoticed task drift. The high-reasoning Codex run predates
that control and is shown only as descriptive context. Each Codex score
contains one execution per task, so no score interval can be estimated from
these data.

\subsection{Agent Layer case study}\label{agent-layer-case-study}

Agent Layer is an open-source orchestration layer for coding agents. It can
optionally include a bundle of editable skills and instructions for planning,
implementation, delegation, and review. The case study evaluates successive
versions of that bundled skill and instruction set on fixed tasks to determine
whether the changes improve score, reduce cost, or both.

The case study holds gpt-5.6-luna at low reasoning, the Codex harness used by
Agent Layer, eight tasks, calibration, and weights fixed. An initial bundle
of custom skills and instructions is followed by
successive revisions. There are ten configurations and 13 completed
evaluations; three configurations were run twice. The endpoint
comparison uses one run per task for the initial and adopted versions.

Recorded token usage is priced at the gpt-5.6-luna rates published
August 16, 2026: \$0.20 per million uncached input tokens, \$0.25 per
million cache-write tokens, \$0.02 per million cached-input tokens, and
\$1.20 per million output tokens. Requests above 272,000 input tokens
use twice the input rate and 1.5 times the output rate
\citep{openai2026luna}.

Across the 13 completed evaluations, the model costs stored with their scored
results total \$27.86 after normalization to the August 16, 2026 rates. Every
case-study cost in this paper is calculated from those evaluation records.

The initial and adopted calibrated scores are compared using the weighted
difference from Section~\ref{retrospective-comparison-of-two-published-arms}.
Because each Codex-harness endpoint contains only one run per task, the
analysis estimates its uncertainty from repeated published Luna medium runs in mini-swe-agent.
Published Luna medium had the closest aggregate score to Codex Luna low on
these tasks. This practical calibration choice does not make the two harnesses
equivalent: the score \emph{p}-value assumes similar task-level repeat-run
variability in both. For cost, all eight task differences favor the adopted
version. A two-sided exact paired sign-flip test measures how unusual that
unanimous result would be if neither version had a consistent cost advantage.

\subsection{Reproducibility artifacts}\label{reproducibility-artifacts}

DeltaSelect is available in Agent Layer at
\url{https://agent-layer.dev/deltaselect}; its source is available at
\url{https://github.com/conn-castle/agent-layer}.
The interactive tool includes the pinned published DeepSWE data used to rank
and calibrate tasks in the published-data analysis. Agent Layer's exported
selections and study reports preserve the task set, weights, model settings,
results, costs, and execution provenance for new studies.
For new studies, request-level billing records count each coordinator and
delegated model call once. Operational incident
logs are retained as provenance but are not part of the results;
verifier build failures and any score-changing correction remain
visible in the evaluated result.

\section{Results and Discussion}\label{results-and-discussion}

Results establish two findings. Published measurements did not transfer
directly to Codex, and repeated use of a fixed eight-task set guided Agent
Layer to a lower-cost skills configuration with a higher observed score.

\subsection{The Codex harness changed the
measurement}\label{the-target-harness-changed-the-measurement}

\begin{figure}
\centering
\includegraphics[width=\linewidth,keepaspectratio,alt={Three plots comparing gpt-5.6-luna scores and costs under published mini-swe-agent and the Codex harness at low, medium, and high reasoning, plus the case-study path.}]{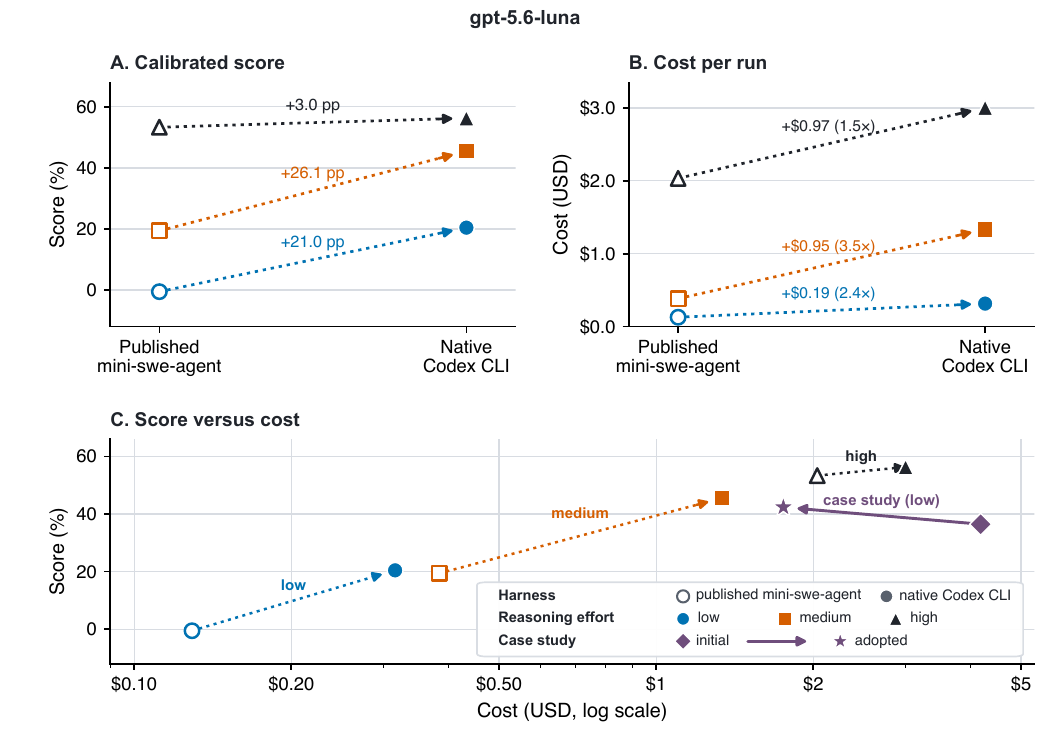}
\caption{Published mini-swe-agent and Codex-harness measurements differ
for the same model and eight tasks. Panels A and B show score and cost
at three reasoning levels; panel C places the case-study path between
the initial and adopted configurations. Each Codex-harness value contains
one run per task, so the figure establishes a planning mismatch rather
than the causal effect of the harness. The high-reasoning Codex-harness
run came from an earlier, non-certified environment.}
\label{fig:harness-comparison}
\end{figure}

Figure~\ref{fig:harness-comparison} compares the estimates used by DeepSWE
with Codex runs of the same eight tasks. At low
reasoning, the published calibrated estimate was -0.54\%; the Codex-harness
run scored 20.43\%, a 21.0-point difference. At medium reasoning, the
published estimate was 19.38\% and the Codex-harness run scored 45.50\%, a
26.1-point difference. At high reasoning, the less-controlled Codex-harness
run was 3.0 points above the published estimate (56.33\% versus 53.34\%).

There is no fixed correction from a published mini-swe-agent result to a
Codex-harness result. The gap was positive at all three measured levels, but
its size changed: the published low-to-medium step was 19.9 points and the
Codex-harness step was 25.1 points. The primary Codex Luna low run scored
20.43\%, and an unchanged repeat scored 14.84\%. Published Luna medium at
19.38\% was closer to both than published Luna low at -0.54\%, which informed
the variance calculation. More generally, the case suggests planning variance
one reasoning level or model tier above the intended execution setting---for
example, Luna medium for Luna low or Opus for Sonnet. The mapping is inexact, but worked as a
practical starting point here. Score, cost, and headroom still have to be
measured in the target harness.

The cost mismatch was operational, not just a price-table discrepancy. At low,
medium, and high reasoning, the DeepSWE estimates were \$0.13, \$0.38, and
\$2.04 for the eight tasks; Codex recorded \$0.32, \$1.34, and \$3.00. The
records do not isolate which harness behavior produced the gap. They do show
that the same model, reasoning level, and tasks cost 1.5--3.5 times the
published estimate. The useful planning number is therefore the first baseline
in the target harness, not a published per-task budget.

The low and medium comparisons hold the model, reasoning level, and tasks
fixed, but each Codex-harness result contains one run per task. They show that
the published score and cost did not carry over; they do not measure how much
of the gap the harness caused. That is enough for the decision at hand: run a
baseline in the target harness before using published numbers to plan a change.

Headroom determined the case-study reasoning level. An earlier six-task Luna
high pilot left five
of six tasks at 96.7--100\% F2P and one at 0\%, leaving almost no room to
measure plausible improvement. Luna low left more failures to fix and cost
less, so the development series used low reasoning. Headroom must be read from
raw F2P rather than the calibrated score, which is unbounded and has no literal
distance from 100\%.

\subsection{Iterating on a frozen task
set}\label{iterating-on-a-frozen-task-set}

\begin{figure}
\centering
\includegraphics[width=\linewidth,keepaspectratio,alt={Chronological plots of calibrated score and total recorded model cost across 13 eight-task evaluations.}]{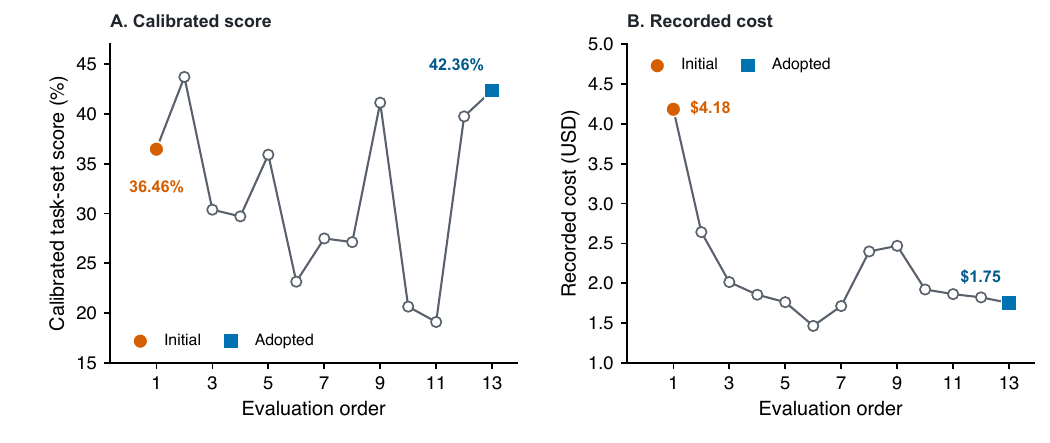}
\caption{Using DeltaSelect during development. Each point is one complete
eight-task evaluation under the fixed design. Panel A shows calibrated score
and panel B shows total recorded model cost. The first and thirteenth points
are the initial and adopted configurations.}
\label{fig:case-study-trajectory}
\end{figure}

Reducing cost was the development goal from the outset. The initial skills
configuration already outscored the bare Codex Luna low baseline on these
tasks, 36.46\% versus 20.43\%. It did so at 13 times the cost: \$4.18
instead of \$0.32, an increase that did not justify the score gain. Across ten
configurations and 13 evaluations, the adopted thirteenth evaluation reached
42.36\% at \$1.75. Raw F2P increased
from 251 to 265 of 411 checks, while recorded model calls fell from 55 to 37.
Figure~\ref{fig:case-study-trajectory} shows the full sequence.

\begin{table}[htbp]
\centering
\small
\caption{Case-study endpoints. The development goal was lower cost without
sacrificing score. The adopted configuration reduced cost on all eight matched
tasks while the calibrated score moved upward, although transferred variance
leaves the score gain uncertain. Dollar figures are total recorded model cost
for the full eight-task evaluation, not cost per task.}
\begin{tabularx}{\linewidth}{@{}>{\raggedright\arraybackslash}Xrrr@{}}
\toprule\noalign{}
Case-study evaluation & Calibrated score & Recorded price (eight tasks) & Raw F2P \\
\midrule\noalign{}
Initial skills and instructions & 36.46\% & \$4.18 & 251/411 \\
Adopted skills and instructions & 42.36\% & \$1.75 & 265/411 \\
Observed change & +5.90 pp & -\$2.43 (-58.1\%) & +14 checks \\
\bottomrule\noalign{}
\end{tabularx}
\end{table}

The score increase was not statistically significant under the transferred
variance model. Propagating published Luna medium task variances through the
fixed coefficients gives standard error 0.05902, 29.23
Welch--Satterthwaite degrees of freedom, and two-sided \emph{p}=0.326. The
standard error is 5.90 points on the percentage scale. The
direction is favorable, but the result does not establish
a population score gain, equivalence, or non-inferiority. The inference also relies
on transferred variance from mini-swe-agent, while each Codex-harness endpoint
contains one run per task.

The strongest endpoint result was cost: all eight matched task costs were lower
under the adopted configuration. The two-sided exact paired sign-flip test gives
\emph{p}=0.0078125, reported as \emph{p}=0.008. Recorded cost fell by \$2.43,
or 58.1\%.

The three repeated configurations moved by 0.7--4.3 calibrated points. Random
variation remains. An unchanged configuration can be repeated when a result is
surprising, but repeats are meant to be a diagnostic exception rather than the
norm; routine replication would defeat the low-cost purpose of the A/B method.

The results in Figure~\ref{fig:harness-comparison} raise an obvious question:
why not just use Luna medium? On the same tasks, the adopted Luna low skills
configuration scored 42.36\% at \$1.75, while bare Luna medium scored 45.50\%
at \$1.34. Medium reasoning is therefore a credible choice for this particular
deployment. The value of skills is broader. They can shape a workflow at any
reasoning level, including levels where headroom and evaluation cost make
improvement difficult to measure, and they let developers specify how an agent
should work rather than accept only the behavior learned during model training.
The practical goal is custom behavior without an unacceptable loss in score or
cost, supporting longer unattended loops and wider adoption.

\subsection{What the iterations
revealed}\label{where-workflow-cost-accumulated}

Successive evaluations on fixed tasks, together with their saved transcripts and
artifacts, made the workflow diagnosable. Simplifying the files helped, but did
not eliminate the expensive control flow. The second evaluation reduced the
run to 43 calls and \$2.64 while increasing score to 43.72\%, but
every task still invoked planning, plan review, and code review. The first
lesson was that orchestration control flow, not prompt-file length, was the
cost driver.

Evaluation 5 tested a bundle of fixes to previously identified defects in
Agent Dispatch, a feature of Agent Layer.
Dispatch lets an agent start a fresh subagent of any supported type in a new
session using that provider's default harness, then wait for or resume that
session. The broader DeltaSelect run history had shown cases where dispatch was silently unavailable or
ignored despite explicit instructions, as well as defects in the wait and
resume functions that spent model calls checking healthy work or failed to
continue unfinished work.

Restricting plan review cut cost to \$2.01 and then \$1.86 in evaluations 3
and 4, but scores fell to 30.38\% and 29.71\%. Fixes to dispatch waits, review
instructions, missing tooling, and recovery lifted evaluation 5 to 35.91\% at \$1.76,
yet its transcripts showed that writing ``once'' did not enforce a global
continuation limit. Evaluations 6 and 7 were cheaper but worse: the allowed
continuation was often consumed before review, leaving no equivalent repair
once review exposed a deeper defect. The unchanged repeat confirmed the
same conclusion despite moving from 23.15\% to 27.49\%: the cheaper workflow
remained much worse than the 35.91\% configuration it replaced.

Evaluation 8 changed the skill so the coordinator resumed the implementer until
it reported completion. That change caused the Scriggo task to go from zero to 46 of 48 checks,
but raised total cost to \$2.40 and did not improve the aggregate. The run also
showed that passing a repository's visible tests did not guarantee passing the
benchmark verifier. Evaluation 9 recovered to 41.13\%, but bundled several
changes, so the improvement cannot be attributed to any one of them.
Evaluations 10--12 tested tighter review and completion policies; scores varied
widely, and evaluation 12 recovered to 39.75\% at \$1.82 while revealing that
reviewer suggestions could expand the work beyond the user's request.

Evaluation 13 addressed that specific failure. It added two scope checks:
plan-review and code-review findings could be accepted only when they remained
within the user's request. The run scored 42.36\% at \$1.75, compared with
39.75\% at \$1.82 in evaluation 12. One run cannot
establish that the two scope checks caused the increase; the transferred-variance
comparison of the initial and adopted endpoints gives \emph{p}=0.326. But the
change directly corrected observed scope expansion, all eight matched task
costs remained below the initial configuration, and the fixed measurement
showed no material regression. The change was therefore adopted.

Requiring statistical significance for every edit would make development
impractical. When a
change corrects a known workflow problem and measurement shows no material
harm, human judgment can justify keeping it. Several directional improvements
can accumulate into an effect large enough to detect at an endpoint even when
no individual step is significant. That judgment is part of the method's value:
the human decides which behavior matters instead of leaving the development
loop fully unattended.

DeltaSelect supplied a stable measurement and a way to inspect why cost and
score moved across the revision process.

\section{Limitations and Future Work}\label{limitations-and-future-work}

DeltaSelect makes one type of repeated development comparison affordable
enough to use. This paper does not show that its task selector is the best
possible selector, that results transfer unchanged across harnesses, or that a
small task set measures broad capability.

\subsection{A small fixed set is intentionally narrow}

A task subset selected to meet a development budget cannot cover every
capability measured by the full benchmark.
Correlation-based selection can omit a rare capability, and calibration cannot
restore omitted content. Repeated development on the same tasks can also
specialize a system to their repositories, graders, or workflow patterns.
Operators should avoid changes that merely teach the agent to pass a selected
task or exploit its verifier. The fixed set supports comparison within a
development cycle, not a claim of broad capability.

\subsection{Small task sets are more sensitive to task quality}

Small sets also magnify bad tasks. OpenAI stopped using SWE-bench Verified after
finding contamination and test problems and withdrew its recommendation of
SWE-Bench Pro after estimating a roughly 30\% broken-task rate
\citep{openai2026signal,openai2026verified}. DeltaSelect cannot repair a broken
benchmark by selecting fewer tasks. Task and environment checksums, readiness
checks, explicit build failures, a predetermined correction registry, and raw
artifacts remain necessary.

\subsection{The selector and scoring choices need direct comparisons}

The experiments do not show that the reliability ranking beats random,
cheapest-first, or median-correlation selection at equal cost. They also do not
isolate the value of fractional scoring, linear calibration, or inverse-error
weighting. Ordinary least squares may fit poorly near the ends of the F2P
range, while related tasks violate the independence assumed by the current
weights and uncertainty calculation. Matched-cost selector comparisons,
nonlinear calibration, equal-weight sensitivity, and covariance-aware methods
remain future work. The task ranking, calibration, and published-data example
use the same DeepSWE snapshot, so the published-data evidence is in-sample and
may be somewhat optimistic.

\subsection{Published results do not transfer directly to deployed systems}

Published benchmark results, such as the DeepSWE trials used here, describe the
model, reasoning level, harness, tools, and environment that produced them.
Practical deployments often change several of those components. Published
trials can still identify candidate tasks and provide initial score, cost, and
headroom estimates, but those estimates will not match another system exactly.
DeltaSelect addresses this gap by requiring a target-harness baseline, but
other harnesses must validate again and may require a different task set.

Each Codex result contains only one run per task, so the case-study endpoint
test uses variance transferred from the closest published configuration rather
than variance estimated in Codex. That transfer keeps the comparison
affordable, but it may misstate uncertainty. Repeated Codex runs would provide
a better variance estimate at greater cost. Several case-study evaluations
also changed more than one instruction, so the 58.1\% cost reduction cannot be
attributed to a single edit.

The evidence comes from one DeepSWE snapshot, one model family, and one custom
harness. The case study shows how DeltaSelect was used; it does not show that
the same score or cost changes will recur elsewhere. DeltaSelect is not a
universal ranking, release benchmark, or safety evaluation.

\section{Conclusion}\label{conclusion}

Frequent A/B decisions are valuable during agent development, but full
coding-agent benchmarks are too broad, expensive, and harness-dependent
to rerun after every change. DeepSWE's repeated public trials show why most
individual tasks are too noisy to trust from one run. DeepSWE runs each task
four times but reduces every result to pass or fail, even when the verifier
records partial progress.

DeltaSelect turns the repeated trials into a fixed, budget-constrained
task set. DeltaSelect conservatively ranks tasks, retains fractional verifier
evidence, maps heterogeneous task results onto a common reference scale,
and allocates the fixed order under an execution budget. The method is not tied
to DeepSWE: the same repeated-trial ranking, budget selection, and calibration
can be applied to new benchmarks for stronger frontier models. The benchmark
and selected tasks can change without changing the method. The result is an
instrument for comparing a baseline with a candidate, not a leaderboard.

In the published DeepSWE example, DeltaSelect estimated that nine selected tasks
and \$113.08 in scored-trial cost would be enough to distinguish Claude Opus 5
medium from Gemini 3.6 Flash high, whose full-benchmark scores differ by 20.3
points. That is 27.1 times less than the full published protocol's
\$3,063.63 recorded cost. Target-harness runs show that published baseline, price, and
headroom cannot be assumed to transfer across harnesses.

DeltaSelect makes iterative measurement affordable, helps diagnose why score
and cost move, and makes uncertainty explicit. In the Agent Layer case study,
recorded evaluation cost fell by 58.1\% while the final optimized
configuration's observed score was higher. If savings of that scale persist in
broader use, DeltaSelect could make benchmark-guided iteration a routine
engineering practice rather than an occasional expensive audit.

\bibliography{references.bib}

\end{document}